\documentclass[a4paper]{article}
\usepackage[top=3cm,bottom=3cm,left=3.2cm,right=3.2cm,headsep=10pt,a4paper]{geometry} 
\usepackage{url} 
\usepackage[english]{babel} 
\usepackage{mathtools,setspace,graphicx,fancyvrb,parskip,lscape,color, rotating,authblk}
\usepackage[protrusion=true]{microtype} 
\usepackage{amsmath,amssymb,amsfonts,amsthm} 
\usepackage[svgnames]{xcolor} 
\usepackage[hang, small,labelfont=bf,up,textfont=it,up]{caption} 
\usepackage{booktabs} 
\usepackage{cite}
\usepackage{tikz}
\usepackage{graphics}
\usepackage{fix-cm}	 
\usepackage{caption,dsfont}
\usepackage{authblk}
\usepackage{tabularx}
\usepackage{afterpage,float}
\usepackage{mdframed}
\usepackage{sectsty} 
\usepackage{adjustbox}
\usepackage{hyperref}
\usepackage{fancyhdr} 
\usepackage{lastpage} 
\usepackage{cancel}
\usepackage[normalem]{ulem}
\usepackage{lettrine} 

\newcommand{\bi}{\begin{itemize}}
\newcommand{\ei}{\end{itemize}}
\newcommand{\be}{\begin{equation}}
\newcommand{\ee}{\end{equation}}

\newcommand{\bc}{\begin{columns}} 
\newcommand{\ec}{\end{columns}}

\definecolor{color1}{HTML}{3BB500}
\definecolor{color1dd}{HTML}{206300} 
\definecolor{color1d}{HTML}{319800} 
\definecolor{color1l}{HTML}{43CF00} 
\definecolor{color1ll}{HTML}{4CEC00} 
\definecolor{color2}{HTML}{5141DA}
\definecolor{color2ll}{HTML}{D6D2FB}
\definecolor{color2l}{HTML}{9A91E9}
\definecolor{color2d}{HTML}{1E0DAC}
\definecolor{color2dd}{HTML}{140B65}
\definecolor{color3}{HTML}{ECB800} 
\definecolor{color3dd}{HTML}{5E4900} 
\definecolor{color3d}{HTML}{AD8700} 
\definecolor{color3l}{HTML}{FFCD1A} 
\definecolor{color3ll}{HTML}{FFE37E}
\definecolor{color4}{HTML}{D4002A} 
\definecolor{color4dd}{HTML}{730017}
\definecolor{color4d}{HTML}{87001B} 
\definecolor{color4l}{HTML}{F2385D} 
\definecolor{color4l}{HTML}{F17C93}

\colorlet{llgreen}{color1ll}
\colorlet{lgreen}{color1l}
\colorlet{dgreen}{color1d}
\colorlet{ddgreen}{color1dd}
\colorlet{dddgreen}{green!40!black}

\usepackage{titling} 

\newcommand{\HorRule}{\color{DarkGoldenrod} \rule{\linewidth}{1pt}} 

\pretitle{\vspace{-30pt} \begin{flushleft} \HorRule \fontsize{30}{30}  \bfseries  \color{ddgreen} \selectfont} 

\title{Non-equilibrium statistical physics, transitory epigenetic landscapes, and cell fate decision dynamics}

\posttitle{\par\end{flushleft}\vskip 0.5em} 

\preauthor{\begin{flushleft}\large \lineskip 0.5em \sffamily \color{ddgreen}} 

\author{Anissa Guillemin$^1$ \&
Michael P.H. Stumpf$^{1,2}$\\
 } 

\postauthor{\footnotesize \usefont{OT1}{phv}{m}{sl} \color{Black} 
$\ ^1$School of BioScience, University of Melbourne, Melbourne, Parkville 3010, VIC, Australia\\
$\ ^2$ School of Mathematics and Statistics, University of Melbourne, Parkville VIC 3010, Australia
\\
 mstumpf@unimelb.edu.au

\par\end{flushleft}\HorRule} 

\date{} 

\begin{document}
  \maketitle
 \thispagestyle{fancy} 

\begin{abstract}
Statistical physics provides a useful perspective for the analysis of many complex systems; it allows us to relate microscopic fluctuations to macroscopic observations. Developmental biology, but also cell biology more generally, are examples where apparently robust behaviour emerges from highly complex and stochastic sub-cellular processes. Here we attempt to make connections between different theoretical perspectives to gain qualitative insights into the types of cell-fate decision making processes that are at the heart of stem cell and developmental biology. We discuss both dynamical systems as well as statistical mechanics perspectives on the classical Waddington or epigenetic landscape. We find that non-equilibrium approaches are required to overcome some of the shortcomings of classical equilibrium statistical thermodynamics or statistical mechanics in order to shed light on biological processes, which, almost by definition, are typically far from equilibrium. 
\end{abstract}

\section{Introduction}
Cells are often seen as the fundamental level from which to start investigating biological systems \cite{Nurse:2011jw}: drill down in detail, or increase resolution, and we end up with intricate molecular processes and arrangements that shape the cells physiology and behaviour. Zoom out, and we observe how cells interact with each other, compete for resources, or self-organise into tissues and whole multicellular organisms \cite{Smadbeck:2016di,Briscoe:2015jsa}. Genes and their gene products---mRNA and proteins---can only fulfil their evolved biological function in the context of cells and organisms \cite{Goodsell:2009aa}; and for complex multicellular organisms we cannot ignore the discrete cellular structure of tissues. 
\par
Cells are themselves highly dynamic and dynamically changing systems. They sense and respond to changes in their environment; they control their internal state, adapting it where necessary; and many cells coordinate their reproduction into pairs of daughter cells. In multi-cellular, but also in some uni-cellular organisms, this process can involve profound changes in the cells' characteristics and states. 
\par
Dynamical systems describe how many natural, including biological, systems change over time. Analysis of ordinary differential equations (ODE) describes the behaviour over time of  {e.g.}, predator prey systems, biochemical reaction systems, neuronal activity, or physiological processes. 
If $x$ denotes the state of the cell (abundances of mRNAs, proteins, lipids etc) then we assume that we have a (high-dimensional) function, $f(x)$, such that the rate of change in $x$ is given by
\begin{equation}
\frac{dx}{dt}=f\left(x\right) 
\label{eq:ODE}
\end{equation}

with ``cell states", $x^*$,  corresponding to the locally stable stationary points of the dynamical system, calculated from
\[
  f(x^*) = 0, 
\]
that is those solutions for which the eigenvalues of the corresponding Jacobian (assuming linear stability analysis suffices), 
\[
J = \left(\frac{\partial f^i(x^*)}{\partial x^j}\right)_{i,j=1,\ldots,d}
\]
 are all less than zero \cite{Jost:2005aa,Kirk:2015kj}. Where linear stability analysis is not sufficient to assess the (local) stability, other, more involved methods need to be invoked. 
\par
The stochastic differential equation (SDE) counterpart to Eq  \eqref{eq:ODE} takes the form
\begin{equation}
dx=f\left(X\right)dt+g\left(X\right)dW_t,
\label{eq:SDE}
\end{equation}
where  $g(x)$ captures the functional form of the stochastic contribution to the dynamics, and $W_t$ is a Wiener Process increment. The stability of stationary points of deterministic dynamical systems under the influence of stochastic dynamics is in general difficult to assess analytically  \cite{Kwon:2005dk,Qian:2010aa}. Instead, in most cases, approximations or simulations are required. 
\par
What we are interested in are new ways of  determining and analysing the functional form of $f(x)$ (and $g(x)$). This would open up the possibility of making more and better mechanistic models in cell biology. With the exception of a handful of simple models of  {e.g.,} embryonic  \cite{Glauche:2011fm,Chickarmane:2012fh,herbach_modelisation_nodate,Zhang:2014bu} and haematopoietic stem cells \cite{Lei:2014fi}, we have precious few mathematical models of the relevant differentiation systems. Analysis of data is thus largely descriptive, but even from these descriptions we can learn or distill some important lessons that could inform mechanistic modelling in the future \cite{maclean_exploring_2018}. Three such examples include molecular noise, the dynamics of gene regulation, and the time it takes for cell-fate decisions to take place. First, noise in gene expression, or cell to cell heterogeneity, appears to be closely associated with the transition between cell states  \cite{gao_universality_2020, moris_histone_2018, guillemin_drugs_2019}. Second, there is clear evidence that the regulation at the gene expression level is highly dynamic and shaped by factors at the epigenetic, transcriptomic, proteomic, and post-translational modification levels  \cite{coulon_eukaryotic_2013, singer_dynamic_2014, bystricky_chromosome_2015, dong_shaping_2017}. We cannot describe this in terms of static gene regulatory networks, and instead need to develop explicitly dynamical descriptions; even then we need to take into account the uncertainty in these networks  \cite{Babtie:2014jg}. Third, the timing of a transition appears to indicate that differentiation is a non-Markovian process  \cite{stumpf_stem_2017}.
\par
In the following we will outline a set of qualitative frameworks for the analysis of cell differentiation dynamics, developing their connections, and follow one of these, \textit{non-equilibrium statistical mechanics}, further in order to characterise transitions between states or cell fates.

\section{Theoretical descriptions of cell fate decisions}
The theory of dynamical systems offers a set of tools that allow us to investigate developmental processes. There is already a set of well studied problems, including different stem cell differentiation systems \cite{Glauche:2011fm,Chickarmane:2012fh,herbach_modelisation_nodate,Zhang:2014bu,Lei:2014fi}, segmentation in insect development \cite{Clark:2016kc}, neural tube formation \cite{PerezCarrasco:2016gn}, and Turing patterns \cite{Maini:2012gq,scholes_comprehensive_2019}. But for the vast majority of systems of concrete biological interest we lack such mechanistic descriptions.
\par
In addition to developing these models from the bottom up, we can also take a more abstract perspective, again grounded in the theory of dynamical systems. Below we outline two such approaches.

\subsection{Qualitative dynamics of cell differentiation}
If we identify cell fates with the stationary points, $x^*$, of a dynamical system, then, even if we do not know the structure and form of the dynamical system,  {i.e.,} we do not know the mathematical form of $f(x)$ in Eq \eqref{eq:ODE} or Eq \eqref{eq:SDE}, we can still make some general qualitative statements. Work in this area, especially by Ren\'e Thom \cite{Thom:1989aa}, was in fact partly inspired by problems in developmental biology.  
\par
\emph{Catastrophe Theory}  \cite{Thom:1989aa,Demazure:2000aa,francois_landscape_2018} was developed to characterise the qualitative behaviour exhibited by dynamical systems as system parameters change. A central tenet of this approach, borne out empirically as well, is that in many cases even high-dimensional systems can be understood in terms of dynamics of a much lower dimensional system. Here the evocative term ``catastrophe" refers to a sudden change in the qualitative nature of the set of solutions of a dynamical system, caused by a smooth or small change in some model parameter, system variable, or control input.  
\begin{figure}
\begin{center}
\begin{tikzpicture}
    \node[anchor=south west,inner sep=0] at (0,0) {\includegraphics[width=0.6\textwidth]{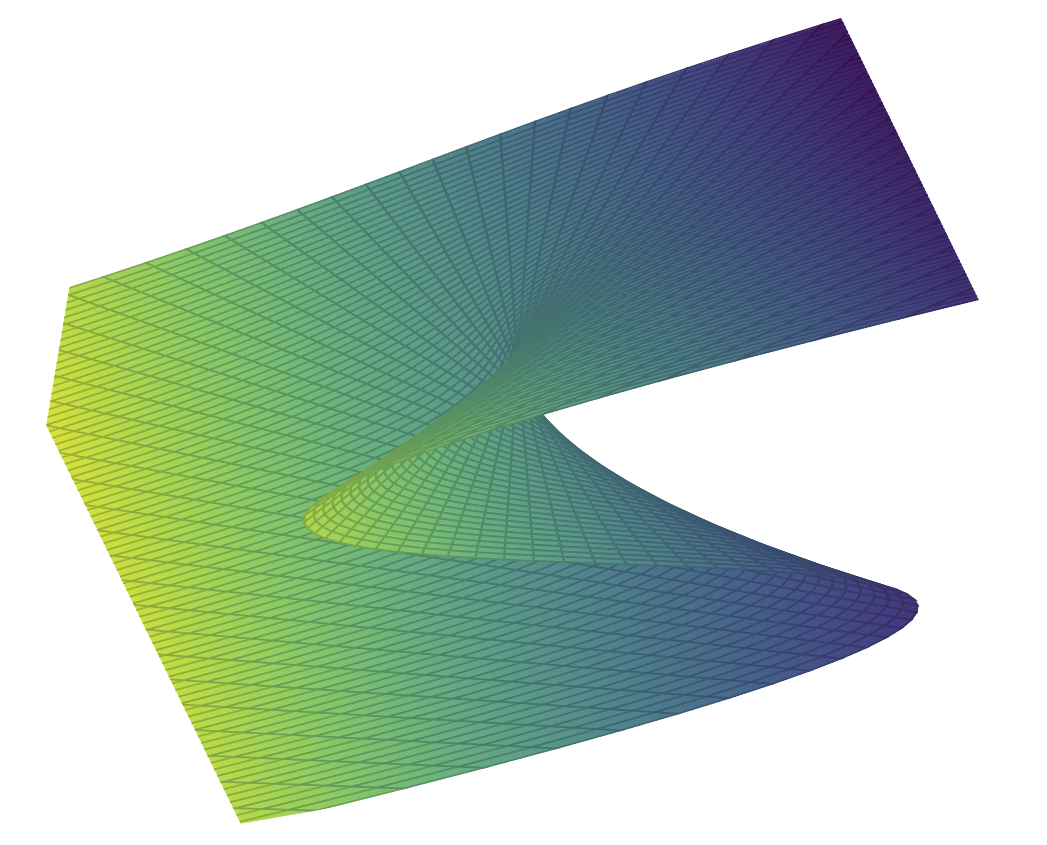}};
    {\node[text centered] (T) at (2.6,7.5) {\Large \sf \textbf{ Cusp Catastrophe}};}

{\node[text centered] (F1) at (6.8,4.8) {\color{white}\sffamily Fate 1}; 
\node[text centered] (F2) at (3,1) {\color{black}\sf Fate 2}; 
\node[text centered] (F3) at (4.,5.2) {\color{gray!20}\sf Stem Cell}; 
\draw[->,ultra thick,gray!90] (8.,5.8) -- node[text width=3.cm,sloped,midway,above] {\color{black}\footnotesize \sf Differentiation} (10.,2.2); 
\draw[<->,ultra thick,gray!90] (4,0.4) -- node[text width=3cm,sloped,midway,below] {\sf \footnotesize Cell-type specific} (9,1.9);}
\end{tikzpicture}
\end{center}
\vspace{0.3cm}
\caption{Example of a Cusp Catastrophe and its relationship to developmental dynamics. The two control variables are a differentiation marker, which leads to loss of stem-like properties, and factors affecting cell-identity, such as the relative abundance of competing transcription factors.}
\label{Figure1}
\end{figure}
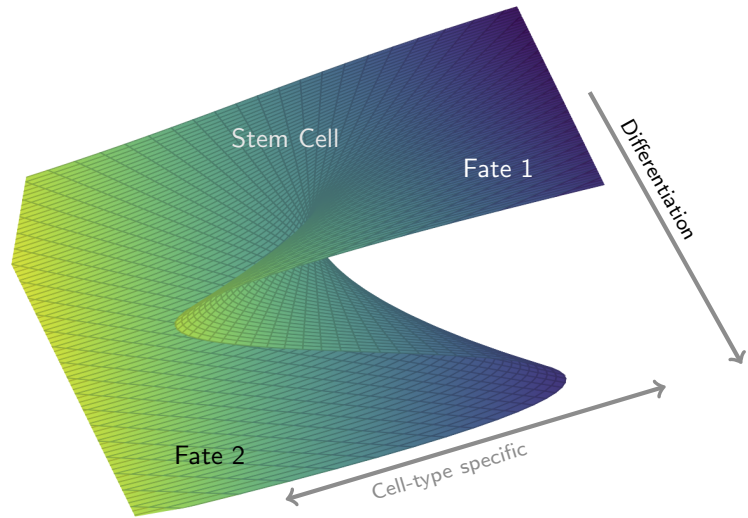
\par
In the example in Figure~\ref{Figure1} such a change is observed in the number of fates that are accessible, as cell-type specific markers are varied. There are regions in parameter space where either one of the fates (Fate 1 or Fate 2) is realised, and regions where both can co-exist. As the differentiation factor is reduced, only a single state, the stem-like state, can exist. In the language of catastrophe theory this is an example of a cusp catastrophe.
\par 
Despite the appeal of this framework there are considerable limitations, including the fact that many of the theoretical results are limited to gradient systems, although stability properties can apply more widely  \cite{zeeman1988}. Certainly a central cornerstone of catastrophe theory, {\em structural stability}  \cite{Demazure:2000aa}, will have wider implications. We mean by this that the qualitative behaviour of a mathematical model or theoretical system remains stable even when the model is changed slightly. Such structural stability of a model  would be desirable, as we know that real-world systems differ from our models, and often quite considerably so. Clearly we should therefore strive in our modelling to only consider structurally stable systems.

\subsection{Models of epigenetic landscapes}
Following Waddington's groundbreaking conceptual work---which has had great influence on Rene Thom's work  \cite{Thom:1989aa}---the well-known epigenetic landscape was long primarily seen as a useful metaphor for developmental processes \cite{conrad_hal_waddington_strategy_1957}. In the 21st century, however, it has increasingly been seen as a computational tool in its own right \cite{moris_transition_2016,huang_cell_2010,Zhou:2012fc,Brackston:2018bb,Brackston:2018kw,wang_potential_2008,Anderson:2019ee}. And there is a rich mathematical literature to draw on, that has only rarely been tapped into so far \cite{francois_landscape_2018}, notably related to Morse theory  \cite{Jost:2005aa,weber_morsewitten_2006}.
\par
Here, however, we follow in the first instance the statistical physics perspective developed above. We shall also restrict ourselves to gradient-like systems,  {i.e.,} where $f(x)$ in Eqs \eqref{eq:ODE} and \eqref{eq:SDE} can be written as
\begin{equation}
f(x) = -\nabla U(x),	
\label{eq:potential}
\end{equation} 
where $U(x)$ is a potential \cite{Jost:2005aa} (or quasi-potential \cite{huang_non-genetic_2009} under suitable circumstances). For stochastic systems (for deterministic systems, \eqref{eq:ODE} resulting densities  are typically sums of Dirac $\delta$ functions), we have for the probability density of the solution of Eq  \eqref{eq:SDE} to be at $xdx$,
\begin{equation}
\pi(x) = \exp(-U(x)),
\end{equation}
details to this can be found in  \cite{Zhou:2012fc,Brackston:2018bb,Brackston:2018kw}.
\par
Local minima of the potential correspond to (locally) stable attractors of the dynamics and can thus be related to cell states  \cite{moris_transition_2016,Ventre:2020aa}. And Morse theory \cite{Jost:2011aa,Demazure:2000aa}, which applies to gradient systems, allows us to relate the different fixed-points of gradient systems in the case of deterministic dynamics. Notably, any set of locally stable fixed points, is separated by (at least) one saddle node,  {i.e.,} a fixed point where the Jacobian of the potential, $U(x)$, has both positive and negative eigenvalues  \cite{Jost:2005aa,Demazure:2000aa}.

\subsection{Statistical mechanics of cell differentiation}
\begin{figure}
\begin{center}
    \includegraphics[scale=1]{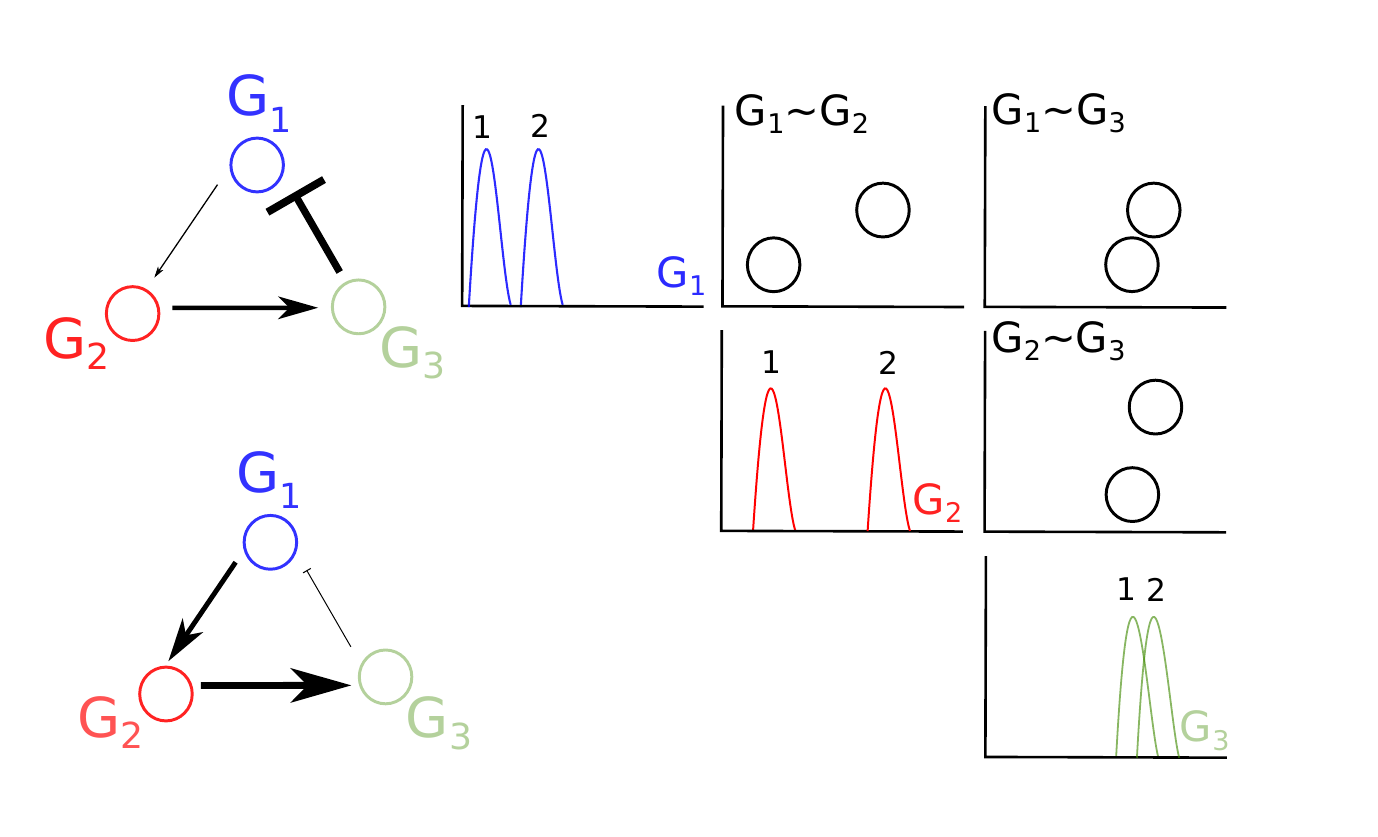}
    \caption{Representation of the definition of macrostates and microstates for the case of a small gene network representing the phenotype. Macrostate A is represented here by the top network; macrostate B by the network at the bottom. The macrostates correspond to distributions over different microstates. The microstates here are interpreted as the (joint)  expression levels of genes, $G_1$ (gene 1), $G_2$ (gene 2) and $G_3$ (gene 3). The expression levels of $G_3$ overlap between the two macrostates, but the $(G_1, G_2, G_3)$ microstates associated with the macrostates are distinct. }
    \label{Figure2}
    \end{center}
\end{figure}

A different approach is rooted in theoretical physics and has found widespread use in  {e.g.,} ecology  \cite{deVladar:2011kz}, signal transduction \cite{mc_mahon_information_2015,Jetka:2018hw,SzymanskaRozek:2019gf}, and gene regulation  \cite{Sasai:2003fe,Benecke:2008}, but it has also been shown to be helpful in developmental and stem-cell biology \cite{garcia-ojalvo_towards_2012,macarthur_statistical_2013,Zhang:2014bu,stumpf_stem_2017}. Statistical physics links microscopic behaviour of  {e.g.,} molecules or atoms, to macroscopic observables such as pressure. In the context of cell biology, the precise molecular composition characterise \emph{microstates}, whereas the cell-types are the observable \emph{macrostates}. Just as in statistical physics, each macrostate is associated with a large number of corresponding microstates. In biological terms, the microstates associated with a given macrostate represent all molecular configurations (chromatin states, mRNA and protein concentrations, transcription factor activities etc) corresponding to a given cell state (Figure~\ref{Figure2}).

\par
Statistical mechanics is concerned with the long-term behaviour of a system \cite{Chandler:1987aa,Attard:2012aa}, in particular assigning probabilities of different macrostates being realised. We will start by defining some of the terminology. First, we use Latin and Greek letters to denote micro- and macrostates, respectively; here and below we follow the exceptionally clear terminology set out by Attard  \cite{Attard:2012aa}. We assume that we can meaningfully define weights for the different microstates, $\omega_i$. We then have for the weights of the corresponding macrostates, $\omega_\alpha$, 
\begin{equation}
\omega_\alpha = \sum_{i\in\alpha} \omega_i.
\end{equation}
Summing over all microstates and macrostates must necessarily give the same value, $W$, that is we have
\begin{equation}
\sum_\alpha \omega_\alpha = \sum_i \omega_i = W,
\end{equation}  
where $W$ is the total weight (to be made precise below). For notational convenience we use the sum symbol, $\sum$, rather than the integral (as would be more appropriate for continuous state spaces).  From the weights we obtain the probabilities of micro- and macrostates,
\begin{equation}
p_i = \frac{\omega_i}{W}, \qquad \text{ and } \qquad p_\alpha = \frac{\omega_\alpha}{W}, 
\end{equation}
with which we can, for example, obtain expectation values for functions,  {e.g.,}  across a macrostate, by averaging 
\[
E[r]_\alpha = \sum_{i\in \alpha} p_i r_i.
\]
Here $r$ could denote cell size, or gene expression level associated with a given cell type, if $\alpha$ refers to cell types. Higher moments, including variances are calculated similarly.
\par
The entropy plays a central role in statistical mechanics, and here is defined as
\begin{equation}
S = \log(W),
\label{eq:totEnt}
\end{equation}
or, if we consider the entropy of a macrostate, as
\begin{equation}
S_\alpha = \log(\omega_\alpha)=\log\left(\sum_{i\in \alpha}\omega_i\right).
\label{eq:EntMacro}
\end{equation} 
If we have a uniform distribution over the microstates we can assign unit weight to each microstate, $\omega_i=1, \forall i$. The entropy of a macrostate is then simply, $S_\alpha = \log n_\alpha$,
where $n_\alpha$ is the number of microstates corresponding to macrostate, $\alpha$. 
\par
One of the central results of statistical mechanics is encapsulated in the {second law of thermodynamics}, which states that entropy never spontaneously decreases  \cite{Chandler:1987aa,Attard:2012aa}. Thus spontaneous change will only ever occur if a change in state leads to an increase in entropy. So in this picture, a change from state $\alpha$ to $\alpha'$ will only be observed if $S_{\alpha'}> S_{\alpha}$; or, in the case of uniform weight over microstates, if $n_{\alpha'}>n_\alpha$. 
\par

\section{Epigenetic landscapes, entropy, and cell fate transitions}
We next link the epigenetic landscape with entropy, first by considering equilibrium statistical mechanics, then by considering transitions between states. We start by developing the total entropy of the system, Eq  \eqref{eq:totEnt}, further,
\begin{align}
S &= \log(W) = \sum_\alpha p_\alpha \log\left(\frac{ \omega_\alpha W}{\omega_\alpha}\right)\\
& = \sum_\alpha p_\alpha S_\alpha - \sum_\alpha p_\alpha\log p_\alpha.
\label{eq:entropy}
\end{align}
Here the second term is the conventional representation, which captures the uncertainty associated with respect to which macrostate, $\alpha$, the system is in. The first term captures the internal uncertainty associated with the macrostates; this is often ignored because in many applications only differences in entropy matter, and the whole term can then be viewed as an additive constant. But, for example, when considering a system coupled to a reservoir, this term does matter profoundly.  It is also important for the case where we consider different macrostates, which below include alternative definitions of cell types.  
\par
\subsection{Macrostates for cell biology}
Identifying macrostates for cell biology is, perhaps surprisingly, non-trivial. Microstates are versions of gene expression states, associated to a macrostate \cite{garcia-ojalvo_towards_2012,macarthur_statistical_2013,stumpf_stem_2017}. Many of these microstates will never be attained \cite{Brackston:2018bb}. We assume that the whole state space can, in principle, be specified and we denote it by $\Gamma$. The whole set of macrostates, called a {\em collective},  has to cover all potential states in a non-overlapping manner, that is
\begin{equation}
\bigcup_\alpha \alpha = \Gamma = \bigcup_i x_i \qquad \qquad \text{ and } \qquad \qquad \alpha \cap \alpha' = \emptyset. 
\end{equation}
The second condition is typically easy to meet, the first is slightly more problematic: we have to assign  microstates that are potentially never observed  \cite{Brackston:2018bb} to appropriate macrostates. We discuss an almost certainly incomplete list of suitable macrostates for cell biology in the following.

\paragraph{Phenotypic Definition:}  If we have a set of objective phenotypic markers (morphology or cell surface marker), $S=\{s_1,s_2,\ldots, s_C\}$,  we may use this as a base from which to define macrostates; $\alpha$. We then, have, however, three types of microstates: (i) microstates that are observed in these cell-types; (ii) microstates that are never observed; these have weight $\omega_i\longrightarrow 0$ and can be ignored without biasing or distorting {e.g.,} the calculation of state-transition probabilities. Finally, (iii) there are microstates that do not conform to these previously defined cell-types and are thus not assigned. These can, however, have finite probability. Their assignment to macrostates is {\em a priori} complicated: they may correspond to new cell-types or sub-types; they may correspond to intermediate cell states \cite{Munoz:2012ip,maclean_exploring_2018}; or they may be fleetingly visited as cells explore the molecular states available \cite{Brackston:2018bb}. These states need to be considered with some care if we want to base macrostates on phenotypic cell definitions.

\paragraph{Data-Driven  Definition:} Alternatively, observed microstates can be subjected to statistical analysis, perhaps, unsupervised learning to group them together and then assign macrostates to clusters  \cite{gao_universality_2020,maclean_exploring_2018,Bergen:2020ju}. The ambiguity of clustering  \cite{Hastie:2009aa}---especially whether to lump small clusters together, or split larger, more extended clusters---is, of course, encountered in this approach. But because of the practical irrelevance of microstates that are never encountered (see above) this approach seems sensible, and unproblematic.   

\paragraph{Dynamical Systems Definition:} We can use ideas from the theory of dynamical systems \cite{Jost:2005aa}. For the deterministic case we can group all microstates, $x_i$ which, for $t\longrightarrow \infty$ go to the same stationary state into the same macrostate. This definition assigns every point in $\Gamma$ to one and only one macrostate. However, generalisation to stochastic dynamics is not straightforward; furthermore, it does not capture the role of saddle node fixed points, which may play an important role in defining intermediate cell states \cite{maclean_exploring_2018}. 

\paragraph{Mixed Macrostates:}
The advantage of the form for the entropy given in Eq  \eqref{eq:entropy} is that we can combine different collectives of macrostates, here denoted by $\alpha$, $\beta$, and  $\gamma$. We can calculate the weight of such mixtures using the usual laws for joint probabilities,  {e.g.,} we have
\[
\omega(\alpha\beta\gamma) = \sum_{i\in \alpha\cap\beta\cap\gamma} \omega_i.
\] 
With this, it becomes possible to combine the macrostate definitions above and overcome their individual limitations. We can also, through simple relabelling of macrostates, simplify the notation and have a single subscript to denote the new ``mixed macrostate".

\subsection{Non-stationary epigenetic landscapes}
One problem related to the difficulty in developing a statistical mechanics for stem cell biology, comes from the fact that much of the appeal of statistical mechanics lies in the fact that entropic arguments can be used to determine (most probable) system states. The second law of thermodynamics, in particular, states that entropy never decreases spontaneously, and that the maximum entropy state is the one realised with high probability \cite{Chandler:1987aa,Attard:2012aa}. If the macrostates are not coherently defined then entropy and $\sum_{i\in\alpha} p_i\log p_i $ cannot be used to assign the most probable states.
\par
Reports, for example, that entropy across cell-populations becomes maximal around the transition state are thus not necessarily in violation of thermodynamics  \cite{gao_e_nodate}: clearly the most probable states ({i.e.,} the states with highest probability as $t\longrightarrow \infty$) will correspond to fixed points and their vicinity. High entropy at transition states could either reflect poor definitions of cell states; or this could simply reflect that the concepts from equilibrium thermodynamics and statistical mechanics are of limited use in this regime.
\par
Both explanations seem eminently plausible, and non-equilibrium statistical mechanics and thermodynamics may offer solutions to the second problems, in particular. We will sketch out two such solutions: a brief introduction to transition probabilities between states and how we can use them to reconcile some of the experimental results. Finally, we briefly turn to considering dynamic epigenetic landscapes.

\subsection{Transition probabilities and entropy of transitions}
In non-equilibrium theories we consider explicit change over time. One convenient way is to consider transitions over time $\tau$, $j \stackrel{\tau}{\longrightarrow} i$, as states of interest. We have for the weight of a microstate $j$
\begin{equation}
    \omega_j = \sum_i \omega(j,i|\tau),
\end{equation}
that is the weight of a microstate, $j$  is equal to the sum over the weights of transitions out of $j$ into any other microstate.  
The weight of a transition between (suitably defined) macrostates $\beta\longrightarrow \alpha$ (where $\alpha$ and $\beta$ can be in the same or different collectives) is then given by
\begin{equation}
    \omega(\beta,\alpha|\tau) = \sum_{i\in\alpha}\sum_{j\in\beta}\omega(j,i|\tau).
    \label{eq:transweight}
\end{equation}
Crucially, even if transitions between microstates were deterministic, transitions between macrostates  will still be stochastic, because specification of macrostates, $\alpha$ and $\beta$, does not precisely define the start and end microstates, $i\in\alpha$ and $j\in \beta$  \cite{Attard:2012aa}. 
\par
In order to normalise the weights, we have to sum over all states, or equivalently, all transitions that can occur, and we get
\begin{align}
    W &= \sum_i \omega_i = \sum_{i,j}\omega(j,i|\tau) \nonumber \\
    &= \sum_\beta \omega_\beta = \sum_{\alpha,\beta}\omega(\beta,\alpha|\tau).
\end{align}
With this we get for the probability of a transition from state $i/\alpha$ to state $j/\beta$ to occur at some time in the future $\tau$, we have
\begin{equation}
    p(j,i|\tau) = \frac{\omega(j,i|\tau) }{W} \qquad \qquad \text{ and } \qquad \qquad p(\beta,\alpha|\tau) = \frac{\omega(\beta,\alpha|\tau) }{W}.
\end{equation}
We then have for the conditional probability of ending up in state $\beta$ given that the system starts in state $\alpha$
\[
p(\beta|\alpha,\tau) = \frac{\omega(\beta,\alpha|\tau) }{\omega(\alpha)}.
\]
Now analogously to Eq \eqref{eq:EntMacro} we can define a new entropy for the transitions between macrostates,
\begin{equation}
    S^{(2)}(\beta,\alpha|\tau) = \log \omega(\beta,\alpha|\tau),
\end{equation}
which we refer to as the second entropy. The advantage of this formalism is that for non-equilibrium systems and equilibrium systems alike, the most likely state transition, given the current state, $\beta$ can be obtained  by determining the state $\hat{\alpha}(\tau|\beta)$ which maximises the second entropy, 
\[
\frac{\partial S^{(2)}(\beta,\alpha|\tau)}{\partial \alpha}\Bigg|_{\alpha=\hat{\alpha}} = 0.
\]
This shifts the focus of the analysis from states to transitions, and may offer a better perspective on cell differentiation than a conventional equilibrium statistical mechanics perspective.

\begin{figure}[t]
\includegraphics[width=\textwidth]{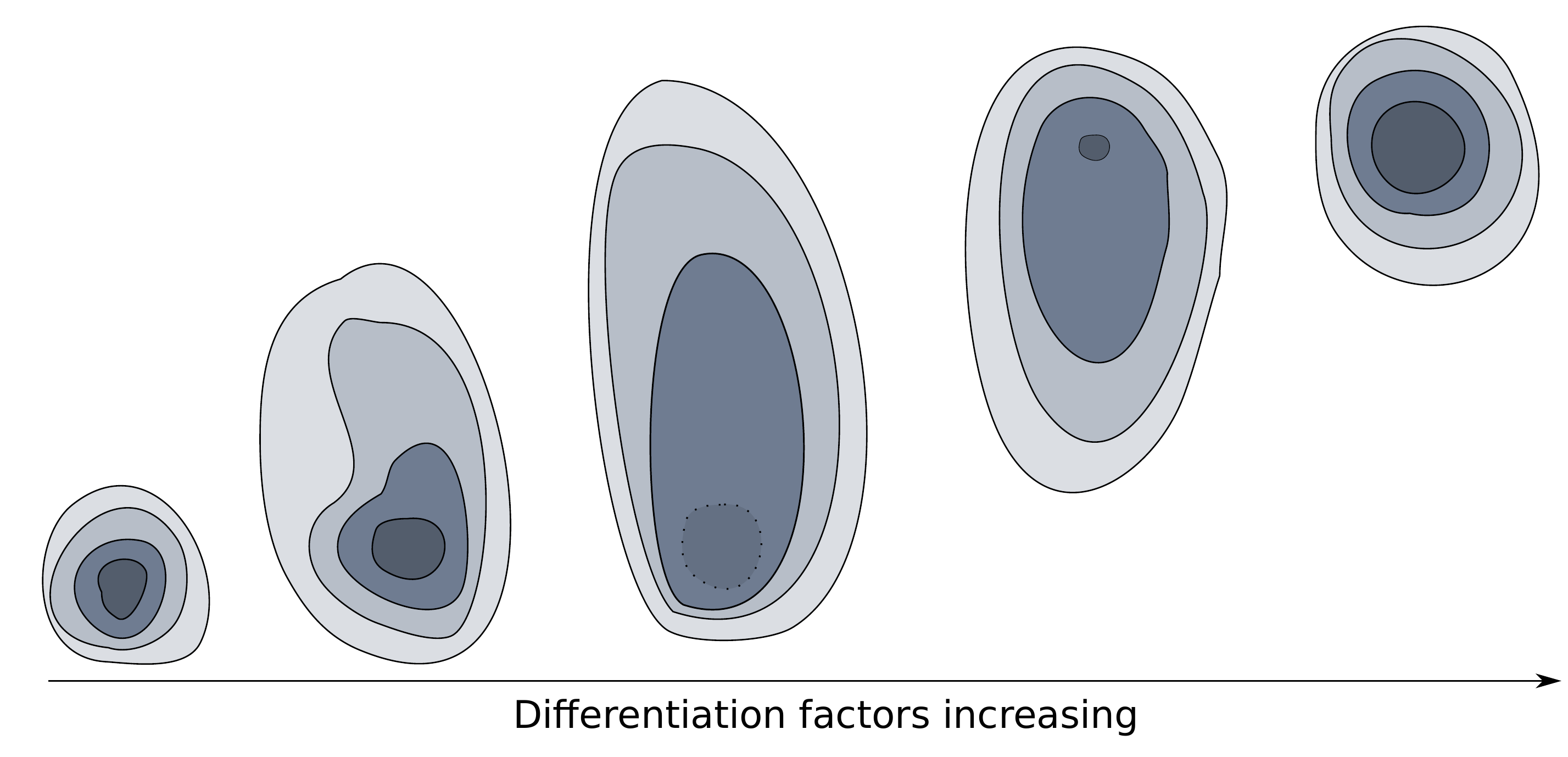}
\caption{The epigenetic landscape, and our quasi-potential representation thereof, will change over time and in response to the cellular environment. This is illustrated here, where the progression along a developmental trajectory leads to the dissolution of old and the creation of new potential minima aka cell types. Whether this evolution of surfaces is necessarily smooth is not clear.  }
\label{Figure3}
\end{figure}

\subsection{Transitory landscapes}
This brings us to the final point, and one that may have been implicitly presaged by Waddington \cite{conrad_hal_waddington_strategy_1957}, and is also apparent in the work of Thom  \cite{Thom:1989aa} and others (see  \cite{francois_landscape_2018} for a recent review): the epigenetic landscape should not be viewed as a fixed object in time, but one which changes dynamically \cite{Brackston:2018bb}. For mathematical models of developmental systems---typically of the essential gene regulation networks---it is possible to calculate the corresponding epigenetic landscapes as quasi-potentials, $U(x)$  \cite{huang_cell_2010,Zhou:2012fc,francois_landscape_2018}. These networks change---some interactions become more prominent, while others fade as  {e.g.,} transcription factor activity changes in response to external signals  \cite{Thorne:2012ia,McMahon:2014cs,Chan:2017cc}---and so do the quasi-potentials, taking the system through a qualitative change point.
\par
In this view (Figure~\ref{Figure3}) the landscape changes in response to signalling and the resulting minima of the quasipotential, $U(x;\zeta)$, which now explicitly depends on an external signal or differentiation factor, shifts position in statespace. In this view we make the stationary states of the system dependent on $\zeta$, that is
\[ 
x^* = x^*(\zeta)
\]
for the $x^*$ that solve
\begin{equation}
\nabla U(x^*;\zeta) = 0.
\end{equation}
\par
The set of solutions, $x^*$, does not need to behave in a continuous manner with the control variable, as is clear, for example, if $\zeta$ induces a bifurcation. {\em A priori } it is not clear if the potential has to vary smoothly.
\par
For a transitory landscape, we could potentially treat the landscape associated with $\zeta$ as the macrostate, but we have to note here that the control variable, $\zeta$, will in practice rarely be scalar: differentiation into a more specialised cell-type typically depends on more than a single molecular factor  \cite{Smith:2017jp}. 

\section{Conclusions}
From a merely conceptual tool the Waddington or epigenetic landscape has been slowly morphing into a framework for the qualitative and quantitative analysis of real biological systems. There are two areas where further investigation and development are likely to bear fruit, and which we discussed above.
\par
First, ideas from dynamical systems theory, Morse theory, catastrophe theory, and especially concepts related to structural stability, have important implications for the mathematical analysis of dynamical systems. 
\par
A crucial challenge to their widespread use is, however, that (i) they are typically restricted to gradient systems; (ii) they are only valid for deterministic systems. There is some reason to be hopeful that analysis of deterministic systems can be meaningful for our understanding of stochastic dynamical systems. But this may require detailed case-by-case analysis, as some hallmarks of deterministic dynamical systems, such as certain types of bifurcations, may not translate to their stochastic counterparts, or \emph{vice versa}.
\par
The second point relates to applying statistical mechanics to  cell differentiation. There is obvious appeal to doing so as has been detailed before. There are two shortcomings to this, however: (i) equilibrium statistical mechanics rests on assumptions that almost certainly do not hold in the context of living and changing systems; (ii) much of the appeal of statistical mechanics stems from the fact that entropic considerations can point towards the state a system will be in. Defining the relevant macrostates is problematic; and translating empirical entropy estimates into  {e.g.,} the likelihood of a given cell-state being obtained, is not possible in the conventional framework. There is, however, as we have sketched out here, some scope to resolve these outstanding issues by adopting a non-equilibrium perspective, and better definitions of cellular macrostates. 
\par
The concept of a transitory landscape \cite{brackston_transition_2018}, may be an attractive way of combining the dynamical systems perspective pioneered by Thom \cite{Thom:1989aa} and others \cite{Siggia:2018vg,francois_landscape_2018}, with the statistical mechanics perspective, especially if an appropriate non-equilibrium framework is used.

\section*{Acknowledgments  }
We would like to thank the members of the {\em Theoretical Systems Biology Group} for many fruitful discussions of cell-fate decision making processes. Anissa Guillemin and Michael P.H. Stumpf gratefully acknowledge funding from the {\em Driving Research Momentum} fund.

\section*{Conflict of interest}
The authors declare that they have no conflict of interest.

\providecommand{\href}[2]{#2}
\providecommand{\arxiv}[1]{\href{http://arxiv.org/abs/#1}{arXiv:#1}}
\providecommand{\url}[1]{\texttt{#1}}
\providecommand{\urlprefix}{URL }

\end{document}